\documentstyle[aps,multicol,prl,epsf]{revtex}
\begin{document}
\draft
\title{Broken unitarity and phase measurements\\ in Aharonov-Bohm interferometers}
\author{O. Entin-Wohlman$^a$, A. Aharony$^a$, Y. Imry$^b$, Y. Levinson$^b$ and
A. Schiller$^c$}
\address{$^a$School of Physics and Astronomy, Raymond and Beverly Sackler
Faculty of Exact Sciences, \\ Tel Aviv University, Tel Aviv 69978,
Israel\\ }
\address{$^b$Department of Condensed Matter Physics, The Weizmann
Institute of Science, Rehovot 76100, Israel}
\address{$^c$Racah Institute of Physics, The Hebrew University, Jerusalem
91904, Israel}

\date{\today}
\maketitle
\begin{abstract}
Aharonov-Bohm mesoscopic solid-state interferometers yield a
conductance which contains a term $\cos(\phi+\beta)$, where $\phi$
relates to the magnetic flux.  Experiments with a quantum dot on
one of the interfering paths aim to relate $\beta$ to the dot's
intrinsic Friedel transmission phase, $\alpha_1$. For closed
systems, which conserve the electron current (unitarity), the
Onsager relation requires that $\beta=0$ or $\pi$. For open
systems, we show that in general $\beta$ depends on the details of
the broken unitarity. Although it gives information on the
resonances of the dot, $\beta$ is generally not equal to
$\alpha_1$. A direct relation between $\beta$ and $\alpha_1$
requires specific ways of opening the system, which are discussed.
\end{abstract}

\begin{multicols}{2}

The wave nature of an electron is reflected e.g. by the complex
amplitude of the wave transmitted through a quantum scatterer.
Under appropriate conditions (discussed below), Aharonov-Bohm
(AB)\cite{1} interferometers may be regarded as analogs of the
double-slit experiment \cite{2,3}, in which the transmission
through two paths is $T=|t_{12}|^2$, with
\begin{equation}
t_{12}=t_1+t_2e^{i\phi}, \label{2slit}
\end{equation}
where $\phi=e \Phi/\hbar c$, $\Phi$ being the magnetic flux
enclosed by the two paths. The path amplitudes $t_i=|t_i|e^{i
\alpha_i}$ may contain the effects of obstacles \cite{4}, e.g. a
quantum dot (QD) on path 1, whose non-trivial (gate voltage
dependent) transmission phase $\alpha_1$ can be influenced by
electronic correlations\cite{5,6}. Assuming the two-slit formula,
Eq. (\ref{2slit}), the Landauer conductance\cite{7} through the
interferometer, $G=(e^2/h)T$, then includes the term\cite{8}
$\cos(\alpha_2-\alpha_1+\phi)$, which is sensitive to the phase
difference. However, in ``closed" or {\it ``unitary"}
interferometers (inside which the electron number is conserved),
time-reversal symmetry implies the Onsager relation\cite{9},
$G(\Phi)=G(-\Phi)$. This relation holds for both finite and
infinite systems. Hence, $T$ must depend on $\phi$ via $\cos\phi$,
with {\it no phase shift}. Here we show that {\it broken
unitarity} does yield a term $\cos(\phi+\beta)$, where $\beta$
depends in general on the rate and on the details of the electron
loss. The universally assumed equality $\beta=\alpha_2-\alpha_1$
requires special ways of opening the system, which we discuss
below (in the context of some of the experiments\cite{5,6,14,15}).
Specifically, we present an exact example in which this relation
does not hold, and then discuss possible conditions under which it
might hold.

We consider solid-state interferometers, with narrow waveguides
for the electron paths, restricted to the mesoscopic scale in
order to retain the coherence of the conduction
electrons\cite{10}.  AB oscillations in $G(\Phi)$ (in spite of
strong impurity scattering), first suggested in Ref.
\onlinecite{11}, were subsequently observed on metallic closed
systems\cite{12} and in semiconducting samples containing QDs near
Coulomb blockade (CB) resonances \cite{4,13}. In these experiments
$G(\Phi)=G(-\Phi)$, as required by the Onsager symmetry. Further
experiments \cite{5,6,14,15} used {\it open systems}, in which
electrons are lost via additional channels which leave the
interferometer, to obtain a non-zero phase shift $\beta$. Assuming
that $\beta=\alpha_2-\alpha_1$, some of the surprising
experimental results were inconsistent with the theoretical
expectations for the phase $\alpha_1$ of the intrinsic
transmission through the QD\cite{16,17,18,19}. Examples include
the phase lapse between consecutive CB resonances\cite{14,15} and
the non-universal phase shifts at the Kondo resonances\cite{5,6}.

While this paper solves specific theoretical models, the results
can be cast in terms of the various energy scales (e.g. decay
widths) characterizing the system. Thus they are much more general
than the models employed. Below we expound the underlying
model-independent physical principles behind these results.

We first consider a single path, and then connect two paths into
an AB interferometer. The QD transmission is typically
\cite{16,17} defined by the geometry in Fig. 1a: a dot D is placed
on a one-dimensional conductor (described below by a tight-binding
model), which models the narrow electronic waveguides (``leads").
An electron wave with amplitude 1 coming from A (or B) generates a
transmitted wave with amplitude $t_1$ (or $t_1'$), and a reflected
wave with amplitude $r_1$ (or $r_1'$). This is described by the $2
\times 2$ scattering matrix, $S_2=\left (\begin{array}{c c}r_1~
t_1'\\ t_1~ r_1'\end{array}\right )$, mapping the two-component
vector of incoming amplitudes onto those of outgoing ones.
Unitarity implies that the determinant of $S_2$ is
$r_1r_1'-t_1t_1'=e^{2i\alpha_1}$, and $\alpha_1$ is defined (for
the specific geometry of Fig. 1a) as the {\it intrinsic Friedel

\vspace{-1.cm}
\begin{figure}
{\leftline\epsfclipon\epsfxsize=2.in\epsfysize=2in\epsfbox{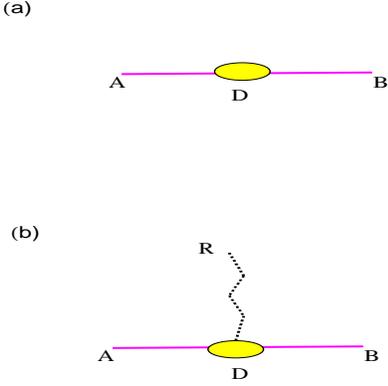}}
\vskip .5truecm \caption{Simple model for a QD (denoted by D)
connected to one-dimensional leads.
(a) A closed system, with no electron losses. (b) An open system,
with a third lead, which connects the QD to a fully absorbing
reservoir R.}
\end{figure}
\noindent phase} \cite{16,17} of the QD. At zero temperature, and
for electrons on the Fermi surface, $\alpha_1$ is equal to the
phase of the Green function on the QD \cite{17},
\begin{equation}
G_D^{(a)}=1/[\epsilon_q-\epsilon_0+e^{iqa}X_{LR}/J],
\label{intrGD}
\end{equation}
where $X_{LR}=J_L^2+J_R^2$, $J_L,~J_R$ represent the
quantum hopping into D from left or right, $J$ is the (tight
binding) hopping between neighboring sites on the leads (with
lattice constant $a$), $\epsilon_q=-2J\cos(qa)$ is the energy
(taken equal to the Fermi energy) of an electron with wave vector
$q$, and $\epsilon_0$ denotes the potential energy on the dot,
determined by the gate voltage. The real parameters $J_L,~J_R$,
and $(\epsilon_q-\epsilon_0)$ may be renormalized by the Coulomb
interactions on the dot, so that $\alpha_1$ contains the effects
of the interactions \cite{kang}. Whenever the gate voltage yields
a resonance, i.e. when the real part of the denominator changes
sign, $\alpha_1$ increases by $\pi$. The width of this jump, given
by the imaginary part of $[G_D^{(a)}]^{-1}$, is determined by
$X_{LR}$, $\Gamma_R=\sin(qa)X_{LR}/J$.

A particular way to break unitarity between A and B is described
in Fig. 1b: a third lead connects the QD to an absorbing electron
reservoir R~\cite{buttiker} (i.e. with a chemical potential which
is slightly lower than that on the emitting source, similar to
that of the absorbing sink).
This QD is described by a unitary $3 \times 3$ scattering matrix
$S_3$, related to the leads from D to A, B and R. However, the $2
\times 2$ matrix $S_2$, which is now a sub-matrix of $S_3$, need
not be unitary! An explicit calculation with such a hopping
Hamiltonian yields that the transmission phase now becomes
$\overline{\alpha_1}$, equal to the phase of the renormalized
Green function, $G_D^{(b)}=1/[(G_D^{(a)})^{-1}-\Sigma]$, where the
complex self-energy $\Sigma$ depends on details of the absorbing
lead. In the simplest case where D is connected to R by the
hopping amplitude $V_1$, we have $\Sigma=-(V_1^2/J)e^{iqa}$. In
particular, its imaginary part, which is proportional to the rate
of electron losses through that lead, contributes to the total
width of the resonance. Thus, the phase $\overline{\alpha_1}$
measured in this case is in general {\it not} the intrinsic
transmission phase of the QD, $\alpha_1$. In fact, for $V_1^2 \gg
X_{LR}$  this contribution of the imaginary self-energy will be
larger than the intrinsic one. It is only when $V_1^2 \ll X_{LR}$
that $\overline{\alpha_1} \approx \alpha_1$. This distinction is
similar to the one obtained in the usual two-slit diffraction
experiment\cite{2} in the following circumstance: inserting an
isotropic resonance scatterer in the upper slit causes the upper
beam to acquire an additional phase shift.
Connecting the source, the scatterer and the screen via a narrow
waveguide produces qualitatively similar results, except that the
width $\Gamma_t$ is now replaced by the typically much smaller
width $\Gamma$ of the resonance against decay into the waveguide.
$\Gamma$ is modified whenever one changes the channels through
which the scatterer can decay.

We next place either Fig. 1a or Fig. 1b as path 1 in the AB
interferometer, as in Fig. 2a or 2b, and calculate

\begin{figure}
\leftline{\epsfclipon\epsfxsize=2.8in\epsfysize=4.5in\epsffile{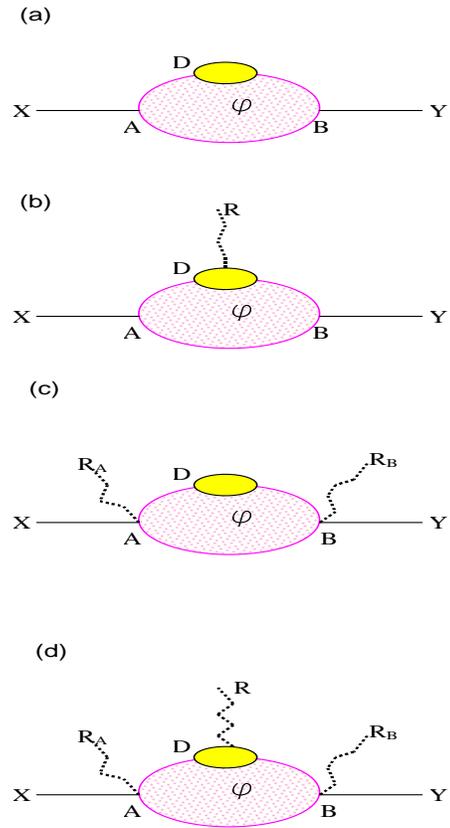}}
\vskip .5truecm \caption{AB interferometers, with a magnetic flux
$\phi$ inside the ring. The text describes calculations of the
transmission amplitude of a wave from terminal X to terminal Y,
for a tight-binding model with single real hopping matrix elements
between A and D ($J_L$), D and B ($J_R$) and on the lower path,
from A to B ($V$). (a) A closed system. (b) Electrons are lost
from the QD via a link to the absorbing reservoir R. (c) Electrons
are lost via links to the absorbing reservoirs R$_A$ and R$_B$.
(d) Same as (c), with the additional loss from D into R.}
\end{figure}
\newpage
\begin{figure}
\leftline{\epsfxsize=2.7in\epsffile{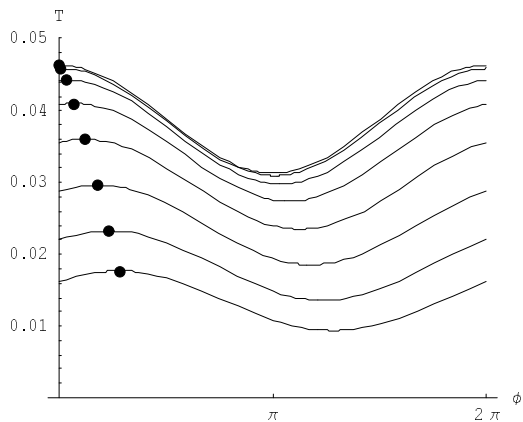}}
\caption{The transmission from X to Y in Fig. 2b, as function of
the AB phase $\phi$;
$J=1,~J_L=J_R=0.1,~V=0.01,~qa=\pi/2,~\epsilon_0=0.1$. The extrema
of the curve shift by the phase $\beta$ (indicated by a point on
each curve), increasing as $V_1$, which measures the rate of
electron losses to the reservoir R, grows from 0 to 0.35 (in steps
of 0.05). Note that the total magnitude of the transmission
decreases with $V_1$, reflecting the same losses.}
\end{figure}
\vskip .5truecm

\noindent the transmission amplitude $t$ for an electron going
from X to Y. For simplicity, we include only one (real, except for
the AB phase $\phi$) hopping matrix element between the sites AD
($J_L$), DB ($J_R$) and AB ($V$). In the unitary case (Fig. 2a),
we find
\begin{equation}
t=CG_D[V(\epsilon_q-\epsilon_0)-J_LJ_Re^{i\phi}], \label{ttt}
\end{equation}
\noindent where $G_D$ is the {\it fully renormalized} Green
function of the dot (containing the effects of all the leads),
\begin{equation}
[G_D]^{-1}=\epsilon_{q}-\epsilon_{0}+ \frac{J X_{LR}
+2VJ_{L}J_{R}\cos\phi e^{iqa}}{J^2e^{-iqa}-V^{2}e^{iqa}},
\end{equation}
and $C=2iJ\sin(qa)/[V^{2}e^{iqa}-J^2e^{-iqa}]$ is a smooth
function of the parameters.

Note that $\alpha_1$ dropped out from the square brackets in Eq.
(\ref{ttt}), which represent the interference: the coefficients
inside the brackets are real, and $T=|t|^2$ depends on $\phi$ only
through $\cos\phi$, as expected from Onsager! $G_D$ does depend on
$\alpha_1$ and on $\phi$, but its dependence on $\phi$ is also
only via $\cos\phi$. The coefficient of $\cos\phi$ in $T$, which
has contributions from both $J_LJ_RV(\epsilon_q-\epsilon_0)$ and
the expansion of $G_D$ in a Fourier series in $\phi$, changes sign
as $\epsilon_0$ increases, yielding a sharp jump of the phase
shift by $\pi$. The vanishing width of this jump is independent of
the dot's intrinsic Friedel phase $\alpha_1$.

We now break unitarity, as in Fig. 2b. Our calculation yields a
similar expression, except that $\epsilon_0$  is now replaced by
the complex $\epsilon_0+\Sigma$. Thus, the absolute value of the
square brackets in Eq. (\ref{ttt}) now contains a term
proportional to $\cos(\beta+\phi)$, with
\begin{equation}
\tan\beta=-{\rm
Im}\Sigma/(\epsilon_q-\epsilon_0-{\rm Re}\Sigma).
\end{equation}
This behavior is portrayed in Fig. 3 (plotted with parameters for
which the dependence of $G_D$ on $\cos\phi$ is weak). Note that
$\beta$ is fully determined by the electron loss into the
reservoir R, and it has {\it no dependence} on the intrinsic QD
transmission phase $\alpha_1$, which follows from Eq.
(\ref{intrGD}). Nevertheless, $\beta$ will change by $\pi$ across
any resonance, where $(\epsilon_q-\epsilon_0-{\rm Re}\Sigma)$
changes sign (up to a shift due to the harmonics of $G_D$). The
width of this change is determined by ${\rm Im}\Sigma$, i.e. by
the rate of electron loss from the QD, and {\it not} by the
intrinsic properties of the dot. In a similar fashion, the phase
shift $\beta$ will exhibit a plateau near $\pi/2$ whenever $|{\rm
Im}\Sigma| \gg |\epsilon_q-\epsilon_0-{\rm Re}\Sigma|$. Such a
plateau is a hallmark\cite{5} of the Kondo effect. However,
establishing its connection to Kondo physics requires more
evidence (such as the enhanced conductance in the CB valley, found
in \cite{5}).

The physical reason for the Onsager symmetry is clear: the
electron wave encircles the interferometer and reflected from the
junctions at A and B many times, complicating the simple two-slit
formula, Eq. (\ref{2slit}). Indeed, our derivation of Eq.
(\ref{ttt}) shows that the cancellation of the phase difference
$\alpha_2-\alpha_1$ from inside the square brackets occurs at each
order in the summation over {\it all} of these reflections. As
already hinted in Ref. \onlinecite{2}, the two-slit formula
requires {\it total absorption} on the junction B (for waves
approaching it from the two paths in the AB ring), thus breaking
unitarity at or before this point. In fact, a sufficient condition
for this formula is that there be {\it no reflections} from
 B backwards to D and A, and similarly from A back
towards D and B. One theoretical way to achieve this is shown in
Fig. 2c: attach to each junction an additional lead to a fully
absorbing reservoir. The full four-link point is now described by
a unitary $4 \times 4$ scattering matrix. One possibility for such
a matrix at point B is
\begin{eqnarray}
S_4=\left (\begin{array}{c c c c} 0 & 0 & cos\omega & -\sin\omega\\
0&0& \sin\omega & \cos\omega\\
\cos\omega & \sin\omega & 0 & 0\\
-\sin\omega & \cos\omega & 0 & 0
\end{array}\right ),
\end{eqnarray}
in which the rows represent R$_B$, Y, A and D. Such a matrix would
arise e.g. for a semi-transparent mirror placed at B, at $45^o$
with the four orthogonal links. Clearly, the $3 \times 3$
sub-matrix corresponding to Y, A, and D is not unitary; however,
its zeroes ensure no reflections back into the ring. Introducing a
similar matrix at A then yields the two-slit formula, Eq.
(\ref{2slit}). Nonetheless, note that the above matrix $S_4$  has
not been derived from a microscopic model (however, similar
elements do exist for microwaves \cite{20}). Such a derivation for
electrons on single-channel leads may require more absorbing leads
(i. e. a larger initial scattering matrix), or more complicated
elements. Furthermore, this matrix has a special and restricted
form, and it is not obvious how to achieve it experimentally.
Finally, the two-slit formula so obtained contains the
transmission amplitudes $t_1$ and $t_2$ of the two individual
paths, and these depend on all the internal details of these
paths, including losses (e.g. as shown in Fig. 2d). The amplitude
$t_1$ will have the desired intrinsic phase $\alpha_1$ of the QD
only when, in addition to the total absorption on junctions A and
B, the width of the dot's resonating state against losses to all
available channels is much smaller than the intrinsic width of the
resonance, $\Gamma_R$.

In real experiments \cite{3,5,14,15}, additional leads are
attached to the ballistic arms of the interferometer, between the
dot and the ``forks" of the interferometer \cite{weiden}. These
leads are ``lossy", as reflected by the small fraction of the
current coming out of the interferometer. When the losses occur
within the back-and-forth reflections of the resonance itself,
then the measured phase will be mainly due to those losses,
similar to our calculations for Fig. 2b. In that case, the AB
phase shift $\beta$ continues to grow with $V_1$, with no
connection to $\alpha_1$. Alternatively, one could have many
weakly coupled absorbing leads along the conducting paths between
the QD and the junctions A or B, outside of this ``rattling"
region. Under appropriate conditions, the reflections from A and B
back into the ring (through the junctions to these leads) become
negligible, the two-slit limit is reached and $\beta$ saturates at
the intrinsic QD transmission phase $\alpha_1$ for a large number
of such leads \cite{new}. Thus, an appropriate specific design of
the unitarity breaking in the experiments should recover the
two-path interference. Considering some of the qualitative results
found in Ref. \onlinecite{15} and in consecutive work, it is quite
possible that these experiments did contain such a design. A
quantitative measurement of the dependence of the measured phase
shift $\beta$ on the strength of the losses could confirm this
possibility.

Two final comments. First, note that in the unitary case, the
interference part of the transmission (square brackets in Eq.
(\ref{ttt})) is {\it real} at zero flux. It may therefore be tuned
to vanish as function of a single control parameter. Such
vanishing may result in a sharp jump of the phase shift measured
in the experiment, from 0 to $\pi$ or {\it vice versa} \cite{JLT}.
This entails the same physical mechanism as the one appearing in
the Fano lineshape \cite{21} (see e.g. Refs. \onlinecite{18,19}
for related suggestions). These considerations may explain some of
the aforementioned experimental puzzles.  Second, unitarity would
also be broken with emitting, rather than absorbing, additional
channels. In view of the lossy experiments, we preferred to
concentrate on the latter.

We thank B. I. Halperin, M. Heiblum, A. Kamenev, Y. Oreg, D.
Sprinzak and H. A. Weidenm\"uller for helpful conversations. This
project was carried out in a center of excellence supported by the
Israel Science Foundation, with additional support from the Albert
Einstein Minerva Center for Theoretical Physics at the Weizmann
Institute of Science and from the German Federal Ministry of
Education and Research (BMBF) within the Framework of the
German-Israeli Project Cooperation (DIP).

\end{multicols}
\end{document}